\documentstyle[pra,aps]{revtex}

\title{Codes for the Quantum Erasure Channel}

\author{M.~Grassl and Th.~Beth}
\address{
	Institut f{\"u}r Algorithmen und Kognitive Systeme,
        Universit{\"a}t Karlsruhe, Am Fasanengarten 5, 
	D--76\,128 Karlsruhe, Germany.
}
\author{T.~Pellizzari}
\address{
        Institut f\"ur Theoretische Physik, 
        Universit{\"a}t Innsbruck, Technikerstra\ss e 25, 
        A--6020 Innsbruck, Austria.}

\begin{document}
\twocolumn
\narrowtext
\maketitle
\begin{abstract}
The quantum erasure channel (QEC) is considered. Codes for the QEC
have to correct for erasures, i.\,e., arbitrary errors at known
positions. We show that four qubits are necessary and sufficient to
encode one qubit and correct one erasure, in contrast to five qubits
for unknown positions. Moreover, a family of quantum codes for the
QEC, the quantum BCH codes, that can be efficiently decoded is
introduced.
\end{abstract}

\section{Introduction}

The prospect of speeding up certain classes of computations by
utilizing the quantum mechanical superposition principle and the
physics of entanglement has received a great deal of attention lately
\cite{reviews}. The potentially useful quantum algorithms so far
include factorization of large numbers \cite{Sh94}, database search
\cite{Grover96} and simulation of quantum mechanical systems
\cite{Lloyd96}. Recent theoretical and experimental progress in atomic
physics and quantum optics has shown that small--scale quantum
computing is feasible \cite{Mo95,Tu95,Ci95,Pe95}.

However, building a quantum computer is an extremely difficult
task. The major obstacle is the coupling of the quantum computer to
the environment which destroys quantum mechanical superpositions very
rapidly. This effect is usually referred to as {\it decoherence}
\cite{Decoherence}. It is thus of crucial importance to find schemes
to actively suppress and {\it undo} the effects of decoherence.

Schemes to protect static quantum states against decoherence were
first found independently by Peter Shor \cite{Sh95} and Andrew Steane
\cite{St95,Ste96}. Their proposals gave rise to a large number of
subsequent publications (see for example \cite{KnLa96,CRSS96} and
references therein). Thus the theory of {\it quantum error--correcting
codes} is increasingly well understood.

In most publications the focus is on finding quantum codes for the
most general error model. These quantum codes can correct for
arbitrary errors at unknown positions in the codeword. However, in
many realistic situations additional information on possible errors is
available. For example, the physical system may permit dephasing
errors only or bit--flip errors only. Of course more efficient codes
are possible for restricted error models.  For example, the smallest
quantum code to correct for errors due to dephasing (or due to
bit--flips) has length three \cite{Br96}. On the other hand, Knill and
Laflamme have shown that the length of the smallest quantum code for
arbitrary errors is five qubits \cite{KnLa96}.

In this paper we consider an error model where the position of the
erroneous qubits is known. In accordance with classical coding theory
we shall call this model the {\it quantum erasure channel} (QEC).
Below a few physical systems are discussed where this model is
applicable. The main results of the present paper are: (i) an explicit
example of a code for the quantum erasure channel (QEC code) with four
qubits which can correct one erasure is presented; (ii) a proof is
presented that four qubits are minimal; (iii) a construction for a
family of QEC codes based on classical BCH codes is given. For these
codes efficient algorithms for correcting erasures exist.

The paper is organized as follows. In Section~\ref{QEC} we introduce
the quantum erasure channel. The error model is discussed and a
physical motivation is given. In Section~\ref{QEC_code} a four--qubit
code for the QEC is given and the proof is presented that four qubits
are minimal.  A construction for quantum BCH codes is given in
Section~\ref{QBCH}.

\section{The Quantum Erasure Channel (QEC)}\label{QEC}

Whenever the position of an error can be determined by an appropriate
measurement the QEC error model applies. In the following we give a
few examples for physical scenarios where this is the case.

\noindent {\bf (i)} If errors are accompanied by the emission of
quanta they can in principle be detected.  For example, if the qubits
are represented by atoms an important source of errors is spontaneous
emission. Spontaneous photons can be observed by photodetection
techniques.  There is, however, the difficulty that spontaneous
photons from free atoms are emitted in a solid angle of $4\pi$ and
will very likely elude observation.  One may circumvent this problem
by modifying the modal structure of the surrounding electromagnetic
field by placing the atoms within a cavity and thereby channeling
spontaneous decay \cite{Bl96}.  Under appropriate conditions photons
escape primarily via cavity decay through the cavity mirrors in a well
defined spatial direction.  There may also be the possibility to
detect the emission of photons by other means, for example via the
photon recoil.  Similarly, if quantum bits are stored in quantized
cavity modes a detected cavity photon indicates an error \cite{Ma96}.

\noindent {\bf (ii)} It is usually assumed that the system space
${\cal H}_{\rm sys}$ is a tensor product of two--dimensional spaces
${\cal H}_2$ (qubits), i.\,e.,
$$
{\cal H}_{\rm sys}={\cal H}_2\otimes\ldots\otimes{\cal H}_2.
$$
However, this is an approximation.  For example, atoms usually have
many levels which may be populated due to an unwanted dynamical
evolution of the system.  Thus the Hilbert space of the system ${\cal
H}_{\rm sys}$ is a tensor product of multi--dimensional spaces with
two--dimensional subspaces used for computing:
\begin{eqnarray*}
 {\cal H}_{\rm sys} &=&{\cal H}_k\otimes\ldots\otimes{\cal H}_k  
\quad\mbox{and}\\
 {\cal H}_{\rm comp}&=&{\cal H}_2\otimes\ldots\otimes{\cal H}_2,
\end{eqnarray*}
where ${\cal H}_{\rm comp}$ is the subspace of allowed computational
states.  Each two--dimensional space ${\cal H}_2$ is a subspace of
${\cal H}_k$, but not necessarily a tensor factor of ${\cal
H}_k$. (For simplicity we assume that the dimension of all tensor
factors is equal.) Therefore, the system space ${\cal H}_{\rm sys}$
can only be decomposed as a direct sum of subspaces
$$
{\cal H}_{\rm sys}={\cal H}_{\rm comp}\oplus{\cal H}_{\rm comp}^{\perp},
$$
and generally not as a tensor product. During error--free computations
the system remains in ${\cal H}_{\rm comp}$.  Any population found in
${\cal H}_{\rm comp}^{\perp}$ is the signature of an error.  Besides,
we can learn about the position of the error by determining which
subsystem has left the allowed Hilbert space ${\cal H}_2$.  The
erroneous subsystem can then be reset by hand to an arbitrary state in
${\cal H}_2$, $|{0}\rangle$ say. As an example we may think of an atom
in which unwanted levels are coupled to the ``allowed'' two--level
system by non--resonant laser interaction. We can measure the
population in these levels for example by applying the quantum jump
technique \cite{q-jumps}.

\noindent {\bf (iii)} QEC codes may be useful in {\it fault tolerant
quantum computing.}  This scheme was recently proposed by Peter Shor
and permits to perform quantum computations and error correction with
a network of erroneous quantum gates \cite{Sh96}.  We may assume that
only quantum gates introduce errors and that errors can be detected by
appropriate measurements.  In this case it is not necessary to use a
quantum code for the most general error model because it is known to
which qubits the quantum gate was applied when an error is
detected. For example, in the cavity QED quantum computer model system
proposed by Pellizzari {\it et al.}~\cite{Pe95} the quantum
information is safely stored in stable Zeeman ground state levels
while no computations are performed. However, during gate operation a
single mode of a quantized cavity is excited, which is much more
fragile a quantum system. A photodetector which records photons
leaking out of the cavity indicates errors in those atoms that are
involved in the current quantum gate.

\noindent {\bf (iv)} It is worthwhile noting that there is a strong
connection of codes for the QEC to the error correction scheme for
quantum gates recently proposed by Cirac {\it et al.}~\cite{Ci96}.
This scheme is designed to correct for a specific but important error
in the ion trap quantum computer during quantum gates. In this error
model errors are caused by decays in the center--of--mass phonon mode
which is temporarily excited during quantum gate operation.  If a
residual population in the phonon mode is found an error is
detected. As above in (iii) the position of the error is known and
thus the QEC error model applies.  In this scheme each logical qubit
is encoded in two physical qubits. One might expect that a four qubit
code is required for this scheme since the smallest code conforming to
the QEC has length four.  However, two qubits are sufficient because
specific assumptions about the type of errors are made.

\section{Codes for the QEC}\label{QEC_code}
\subsection{Conditions on Codes for the QEC}
For the general case, Knill and Laflamme \cite{KnLa96} derived
necessary and sufficient conditions on quantum error--correcting
codes ${\cal QC}$. Given a set of error operators $\{A_i\}$ the
conditions on states $|{c_k}\rangle\in{\cal QC}$ are
\begin{eqnarray}
\langle{c_k}|{A_i^{\dagger}A_j}|{c_k}\rangle
&=&\langle{c_l}|{A_i^{\dagger}A_j}|{c_l}\rangle\label{genExpectCond}\\
\langle{c_k}|{A_i^{\dagger}A_j}|{c_l}\rangle
&=&0
\hphantom{\langle{c_k}|{A_i^{\dagger}A_j}|{c_l}\rangle}
\mbox{for $\langle c_k | c_l\rangle=0$.}\label{genOrthCond}
\end{eqnarray}
For a code of length $N$ that can correct $t$ errors the error
operators $\{A_i\}$ are of a special form. They are all {\em
$t$--error operators}, i.\,e., operators that differ on at most $t$ of
the tensor factors of ${\cal H}={\cal H}_2^{\otimes N}$ from
identity. In (\ref{genExpectCond}) and (\ref{genOrthCond}) it is
sufficient to consider algebra bases for $t$--error operators. The
bases might be tensor products of local bases, e.\,g., the identity
$\openone$ and the Pauli spin matrices
$\{\sigma_x,\sigma_y,\sigma_z\}$, or the operators
$|{0}\rangle\langle{0}|$, $|{1}\rangle\langle{0}|$,
$|{0}\rangle\langle{1}|$, and $|{1}\rangle\langle{1}|$.  In this
paper, we consider the one--error operators $P_{ij}^{k}$ that are the
operators $|{i}\rangle\langle{j}|$ applied to the $k$--th qubit.

For the QEC there are similar conditions. Since the positions of the
errors are known by definition there is no need to separate the spaces
corresponding to errors at different positions. Therefore, in
(\ref{genExpectCond}) and (\ref{genOrthCond}) only $t$--error
operators $A_i$ and $A_j$ that differ from identity at the same
positions have to be considered. But the product of such $t$--error
operators is also a $t$--error operator and can be written as linear
combination of the $A_i$ since they are an algebra basis. Hence,
(\ref{genExpectCond}) and (\ref{genOrthCond}) reduce to
\begin{eqnarray}
\langle{c_k}|{A_i}|{c_k}\rangle
&=&\langle{c_l}|{A_i}|{c_l}\rangle\label{expectCOND}\\
\langle{c_k}|{A_i}|{c_l}\rangle&=&0
\hphantom{\langle{c_l}|{A_i}|{c_l}\rangle}
\mbox{for $\langle{c_k}|{c_l}\rangle=0$.}\label{orthCOND}
\end{eqnarray}
Equations (\ref{genExpectCond}) and (\ref{genOrthCond}) for $t$--error
operators $A_i$ imply equation (\ref{expectCOND}) and (\ref{orthCOND})
for $2t$--error operators since the operators $A_i^{\dagger}A_j$ are
bases for $2t$--error operators. Hence, a quantum error--correcting
code correcting $t$ errors is a $2t$ erasure--correcting code.
 
\subsection{QEC Code with 4 Qubits}
For the general situation it was shown that the shortest code to
encode one qubit and to correct one error has length five
\cite{KnLa96,5bit}. To encode one qubit and correct one erasure,
however, four qubits are sufficient as demonstrated by the code ${\cal
QC}$ given by
\begin{eqnarray*}
   |{\underline{0}}\rangle & = & |{ 0000 }\rangle + |{ 1111 }\rangle\\
   |{\underline{1}}\rangle & = & |{ 1001 }\rangle + |{ 0110 }\rangle.
\end{eqnarray*}
(To simplify the notation, normalization factors are omitted here and
in the remainder of the paper.) In \cite{St95,Ste96} it is shown that
it is sufficient to correct bit--flips in two bases that are Hadamard
transforms of each other. The Hadamard transform of the code ${\cal
QC}$ corresponds to the ``dual'' code ${\cal QC}^{\perp}$ given by
\begin{eqnarray*}
|{\underline{0}^{\perp}}\rangle 
  = H|{\underline{0}}\rangle & = &
  \phantom{+}|{0000}\rangle +|{0011}\rangle 
               +|{0101}\rangle +|{0110}\rangle\\
&&             +|{1001}\rangle +|{1010}\rangle 
               +|{1100}\rangle +|{1111}\rangle\\
|{\underline{1}^{\perp}}\rangle 
  =  H|{\underline{1}}\rangle  & = &
  \phantom{+}|{0000}\rangle -|{0011}\rangle 
              -|{0101}\rangle +|{0110}\rangle\\
&&            +|{1001}\rangle -|{1010}\rangle 
              -|{1100}\rangle +|{1111}\rangle .
\end{eqnarray*}
By definition of an erasure the position of the error is known, but it
is not known what the error is. Since all states of both the code
${\cal QC}$ and its ``dual'' ${\cal QC}^{\perp}$ have even weight, for
both bases a single bit--flip error can be detected by computing the
overall parity. Odd parity indicates an error. Thus, any one--bit
error can be corrected since correcting single bit--flips in both
bases is sufficient.

The code ${\cal QC}$ can be extended by the following two states
$|{\underline{2}}\rangle$ and $|{\underline{3}}\rangle$
\begin{eqnarray*}
   |{\underline{2}}\rangle & = & |{ 1100 }\rangle + |{ 0011 }\rangle\\
   |{\underline{3}}\rangle & = & |{ 1010 }\rangle + |{ 0101 }\rangle
\end{eqnarray*}
with 
\begin{eqnarray*}
|{\underline{2}^{\perp}}\rangle 
  = H|{\underline{2}}\rangle & = &
  \phantom{+}|{0000}\rangle +|{0011}\rangle 
               -|{0101}\rangle -|{0110}\rangle\\
&&             -|{1001}\rangle -|{1010}\rangle 
               +|{1100}\rangle +|{1111}\rangle\\
|{\underline{3}^{\perp}}\rangle 
  =  H|{\underline{3}}\rangle  & = &
  \phantom{+}|{0000}\rangle -|{0011}\rangle 
              +|{0101}\rangle -|{0110}\rangle\\
&&            -|{1001}\rangle +|{1010}\rangle 
              -|{1100}\rangle +|{1111}\rangle.
\end{eqnarray*}
Thus, the extended code encodes not only one, but two qubits and
corrects for one erasure. Note that this code is equivalent to the
code used for error detection in \cite{VGW96}. The existence of a code
with these parameters was shown e.\,g. in \cite{CRSS96}.

\subsection{There is no QEC Code with less than 4 Qubits}
In this section we prove that at least four qubits are required for a
code that can correct one erasure and encodes one qubit. First we
investigate when a quantum code can be shortened.

\noindent{\bf Theorem 1}
{\em 
Let ${\cal QC}$ be a quantum error--correcting code that can correct at
least one erasure. If a one--qubit state $|{\theta_0}\rangle$ is a
factor of a state $|{\phi_0}\rangle\in{\cal QC}$ it is a factor of all
states $|{\phi}\rangle\in{\cal QC}$.
}

\noindent{\bf Proof:\ } Assume w.\,l.\,o.\,g.\  that the first qubit
is a factor, i.\,e., $|{\phi_0}\rangle =|{\theta_0}\rangle
|{\psi_{0}}\rangle$. Inserting the local operator
$P_{|{\theta_0}\rangle}
=|{\theta_0}\rangle\langle{\theta_0}|\otimes\openone$ in
(\ref{expectCOND}) yields for any state $|{\phi}\rangle\in{\cal QC}$
$$
\langle{\phi}|{P_{|{\theta_0}\rangle}}|
   {\phi}\rangle
=\langle{\phi_0}|{P_{|{\theta_0}\rangle}}|
   {\phi_0}\rangle
=1.
$$
Hence, $|{\theta_0}\rangle$ is a factor of every code state.

Thus, we have the following corollary.

\noindent{\bf Corollary 2}
{\em 
If a quantum code ${\cal QC}$ of length $N$ has a one--qubit factor
deleting this position yields a quantum code ${\cal QC}'$ of length
$N-1$ and equal dimension with same error--correcting capabilities.
}

Next, we show that every two--dimensional subspace of ${\cal
H}_2\otimes{\cal H}_2$ contains at least one product state.

\noindent{\bf Lemma 3}
{\em 
For every two--dimensional subspace of ${\cal H}_2\otimes{\cal H}_2$
there is a basis that contains at least one product state, i.\,e., a
state $|{\pi}\rangle =|{\pi_1}\rangle |{\pi_2}\rangle$.
}

\noindent{\bf Proof:\ } Let the subspace be generated by
$\{|{b_1}\rangle ,|{b_2}\rangle\}$. A product state
$|\pi\rangle\in{\cal H}_2\otimes{\cal H}_2$ is characterized by
\begin{equation}\label{prodstate}
\langle{00}|{\pi}\rangle\langle{11}|{\pi}\rangle=
\langle{01}|{\pi}\rangle\langle{10}|{\pi}\rangle.
\end{equation}
Inserting $|{\pi}\rangle =\eta_1|{b_1}\rangle +\eta_2|{b_2}\rangle$ in
(\ref{prodstate}) yields a quadratic equation for the complex
coefficients $\eta_1$ and $\eta_2$:
\begin{equation}\label{quadeq}
0=c_1\eta_1^2+c_{12}\eta_1\eta_2+c_2\eta_2^2
\end{equation}
with
\begin{eqnarray*}
c_1&=&  \langle{00}|{b_1}\rangle\langle{11}|{b_1}\rangle-
        \langle{01}|{b_1}\rangle\langle{10}|{b_1}\rangle\\
c_{12}&=&  \langle{00}|{b_1}\rangle\langle{11}|{b_2}\rangle+
           \langle{11}|{b_1}\rangle\langle{00}|{b_2}\rangle\\
&&\quad     -\langle{01}|{b_1}\rangle\langle{10}|{b_2}\rangle-
              \langle{10}|{b_1}\rangle\langle{01}|{b_2}\rangle\\
c_2&=&  \langle{00}|{b_2}\rangle\langle{11}|{b_2}\rangle-
        \langle{01}|{b_2}\rangle\langle{10}|{b_2}\rangle.
\end{eqnarray*}
If $c_1$ vanishes $|{b_1}\rangle$ is a product state and the lemma
holds. Similarly, $|{b_2}\rangle$ is a product state if $c_2=0$. Now
consider the case $c_1\neq0$ and $c_2\neq0$. The solutions of
(\ref{quadeq}) are given by
$$
\eta_1=\frac{-c_{12}\pm\sqrt{c_{12}^2-4c_1c_2}}{2c_1}\eta_2.
$$
For $c_1\neq0$ and $c_2\neq0$ there is at least one non--trivial
solution with $\eta_1\neq0$ and $\eta_2\neq0$ and thus a product state
exists.

Using Lemma~3 we are able to prove the following theorem.

\noindent{\bf Theorem 4}
{\em 
There is no quantum error--correcting code of length two that can
correct one erasure and encodes one qubit.
}

\noindent{\bf Proof:} Assume that such a code exists. The states
$|{\underline{0}}\rangle$ and $|{\underline{1}}\rangle$ span a
two--dimensional subspace ${\cal QC}$ of ${\cal H}_2\otimes{\cal
H}_2$. According to Lemma~3, ${\cal QC}$ contains a product state
$|{\pi_1}\rangle |{\pi_2}\rangle$. From Theorem~1 follows that both
$|{\pi_1}\rangle$ and $|{\pi_2}\rangle$ are factors of all code states
and thus the code cannot be two--dimensional.

\noindent{\bf Theorem 5}
{\em There is no quantum error--correcting code of length three that
can correct one erasure and encodes one qubit.
}

\noindent{\bf Proof:} Assume that such a code exists. Since there is
no code of length two the states in the code cannot be factored. With
reference to the first qubit the encoding can be written as
\begin{eqnarray*}
|{\underline{0}}\rangle  
  &=& |{0}\rangle |{\Phi_0}\rangle + |{1}\rangle |{\Phi_1}\rangle\\
|{\underline{1}}\rangle  
  &=& |{0}\rangle |{\Theta_0}\rangle  + |{1}\rangle |{\Theta_1}\rangle,
\end{eqnarray*}
where $|{\Phi_i}\rangle$, $|{\Theta_j}\rangle$ are, in general,
unnormalized and non--orthogonal states. The states $|{\Phi_0}\rangle$
and $|{\Phi_1}\rangle$ have to be linearly independent since otherwise
$|{\underline{0}}\rangle$ is a product state and a code of length one
exists (cf.~Corollary~2). Similarly, $|{\Theta_0}\rangle$ and
$|{\Theta_1}\rangle$ have to be linearly independent.

For the projections
$P_{ij}^{(1)}
=|{i}\rangle\langle{j}|\otimes\openone\otimes\openone$, 
$i,j\in\{0,1\}$ equation (\ref{orthCOND}) implies
\begin{eqnarray*}
&&\langle{\underline{1}}|{P_{00}^{(1)}}|{\underline{0}}\rangle
   =\langle{\Theta_0}|{\Phi_0}\rangle=0\\
&&
\langle{\underline{1}}|{P_{10}^{(1)}}|{\underline{0}}\rangle
  =\langle{\Theta_1}|{\Phi_0}\rangle=0\\
&&\langle{\underline{1}}|{P_{01}^{(1)}}|{\underline{0}}\rangle
    =\langle{\Theta_0}|{\Phi_1}\rangle=0\\
&&\langle{\underline{1}}|{P_{11}^{(1)}}|{\underline{0}}\rangle
    =\langle{\Theta_1}|{\Phi_1}\rangle=0.
\end{eqnarray*}
Thus, the subspaces ${\cal H}_{|{\underline{0}}\rangle}$ spanned by
$\{|{\Phi_0}\rangle ,|{\Phi_1}\rangle\}$ and ${\cal
H}_{|{\underline{1}}\rangle}$ spanned by $\{|{\Theta_0}\rangle
,|{\Theta_1}\rangle\}$ are two--dimensional and orthogonal. This
yields a decomposition of the joint Hilbert space of the second and
third qubit:
$$
{\cal H}_2\otimes{\cal H}_2
  ={\cal H}_{|{\underline{0}}\rangle}\oplus
         {\cal }H_{|{\underline{1}}\rangle}.
$$
We now choose an orthonormal basis $B=\{|{b_1}\rangle ,|{b_2}\rangle
,|{b_3}\rangle ,|{b_4}\rangle\}$ for the Hilbert space of the second
and third qubit such that $\{|{b_1}\rangle ,|{b_2}\rangle\}$ and
$\{|{b_3}\rangle ,|{b_4}\rangle\}$ span ${\cal
H}_{|{\underline{0}}\rangle}$ and ${\cal H}_{|{\underline{1}}\rangle
}$, respectively. In the orthonormal basis $B$, the codewords can be
written as
\begin{eqnarray*}
|{\underline{0}}\rangle  
  &=& |{\alpha}\rangle |{b_1}\rangle  
    + |{\beta}\rangle |{b_2}\rangle\\
|{\underline{1}}\rangle  
  &=& |{\gamma}\rangle |{b_3}\rangle  
    + |{\delta}\rangle |{b_4}\rangle,
\end{eqnarray*}
where $|{\alpha}\rangle$, $|{\beta}\rangle$, $|{\gamma}\rangle$, and
$|{\delta}\rangle$ are, in general, unnormalized and non--orthogonal
states in the Hilbert space of the first qubit.

According to Lemma~3, w.\,l.\,o.\,g.\  $|{b_1}\rangle$ and
$|{b_3}\rangle$ can be assumed to be product states. Since a local
unitary transformation of a ${\cal QC}$ yields another ${\cal QC}$
with same parameters, w.\,l.\,o.\,g.\  $|{b_1}\rangle =|{00}\rangle$
can be chosen. At least one factor of the product state
$|{b_3}\rangle$, w.\,l.\,o.\,g.\  the first one, has to be
$|{1}\rangle$ since $\langle{b_1}|{b_3}\rangle=0$.

Therefore, the orthonormal basis $B$ has the form
\begin{eqnarray*}
|{b_1}\rangle  &=& |{00}\rangle  \\
|{b_2}\rangle  &=& k_1|{01}\rangle 
             -k_2{b_{31}^*}|{10}\rangle 
             +k_2{b_{30}^*}|{11}\rangle\\
|{b_3}\rangle  &=& b_{30}|{10}\rangle +b_{31}|{11}\rangle\\
|{b_4}\rangle  &=& -{k_2^*}|{01}\rangle 
              -{k_1^*}{b_{31}^*}|{10}\rangle 
              +{k_1^*}{b_{30}^*}|{11}\rangle 
\end{eqnarray*}
with $|k_1|^2+|k_2|^2=1$ and $|b_{30}|^2+|b_{31}|^2=1$. The codewords
are of the form
\begin{eqnarray}
 |{\underline{0}}\rangle  
   &=& |{\alpha}\rangle |{00}\rangle  
    	 +k_1|{\beta}\rangle |{01}\rangle 
      	 -k_2{b_{31}^*}|{\beta}\rangle |{10}\rangle\nonumber\\
&&\quad	 +k_2{b_{30}^*}|{\beta}\rangle |{11}\rangle\nonumber\\
 |{\underline{1}}\rangle  
   &=&   -{k_2^*}|{\delta}\rangle |{01}\rangle 
     	 +b_{30}|{\gamma}\rangle |{10}\rangle 
         -{k_1^*}{b_{31}^*}|{\delta}\rangle |{10}\rangle\nonumber\\
&&\quad  +b_{31}|{\gamma}\rangle |{11}\rangle 
         +{k_1^*}{b_{30}^*}|{\delta}\rangle |{11}\rangle.
  \label{codeStates}
\end{eqnarray}
If $k_2=0$ the state $|{\underline{0}}\rangle$ would have the factor
$|{0}\rangle$ at the second position and a code of length two would
exist. Therefore we have $k_2\ne0$.

  From equations (\ref{orthCOND}) and (\ref{codeStates}) we obtain the
following conditions:
\begin{eqnarray}
0&=&\langle{\underline{1}}|
      {P_{00}^{(2)}}|{\underline{0}}\rangle
= 
  -k_1k_2\langle{\delta}|{\beta}\rangle\label{gl1}\\
0&=&\langle{\underline{1}}|
      {P_{01}^{(3)}}|{\underline{0}}\rangle
= 
 k_2{(b_{30}^*)^2}\langle{\gamma}|{\beta}\rangle
 -k_1 k_2{b_{30}^*} b_{31}\langle{\delta}|{\beta}\rangle\label{gl7}\\
0&=&\langle{\underline{1}}|
       {P_{10}^{(3)}}|{\underline{0}}\rangle\nonumber\\
&=& 
 -k_2\langle{\delta}|{\alpha}\rangle
 -k_2{(b_{31}^*)^2}\langle{\gamma}|{\beta}\rangle
 -k_1 k_2 b_{30}{b_{31}^*}\langle{\delta}|{\beta}\rangle\label{gl6}\\
0&=&\langle{\underline{1}}|{P_{10}^{(2)}}|
        {\underline{0}}\rangle\nonumber\\
&=& 
 {b_{30}^*}\langle{\gamma}|{\alpha}\rangle
 -k_1 b_{31}\langle{\delta}|{\alpha}\rangle
 +k_1{b_{31}^*}\langle{\gamma}|{\beta}\rangle\nonumber\\
&&\quad
 +k_1^2 b_{30}\langle{\delta}|{\beta}\rangle\label{gl2}\\
0&=&\langle{\underline{1}}|{P_{01}^{(2)}}|
        {\underline{0}}\rangle
= 
  -k_2^2{b_{30}^*}\langle{\delta}|{\beta}\rangle\label{gl3}
\end{eqnarray}
In the sequel we distinguish whether $k_1$ and $b_{30}$ vanish or not:

{
\begin{enumerate}
\item $k_1\ne0$, $b_{30}\ne0$:

  From (\ref{gl1}) follows $\langle{\delta}|{\beta}\rangle=0$ and thus
(\ref{gl7}) reduces to $\langle{\gamma}|{\beta}\rangle=0$.
\item $k_1\ne0$, $b_{30}=0$:

  From (\ref{gl1}) follows $\langle{\delta}|{\beta}\rangle=0$. Equations
(\ref{gl6}) and (\ref{gl2}) reduce to
\begin{eqnarray*}
-k_2\langle{\delta}|{\alpha}\rangle
 -k_2{(b_{31}^*)^2}\langle{\gamma}|{\beta}\rangle &=&0\\
-k_1 b_{31}\langle{\delta}|{\alpha}\rangle
 +k_1{b_{31}^*}\langle{\gamma}|{\beta}\rangle&=&0.
\end{eqnarray*}
This implies $\langle{\delta}|{\alpha}\rangle=0$ and
$\langle{\gamma}|{\beta}\rangle=0$.
\item $k_1=0$, $b_{30}\ne0$:

  From (\ref{gl3}) and (\ref{gl7}) follows
$\langle{\delta}|{\beta}\rangle=0$ and
$\langle{\gamma}|{\beta}\rangle=0$.
\item $k_1=0$, $b_{30}=0$:

The basis states $|{b_3}\rangle$ and $|{b_4}\rangle$ are
$|{11}\rangle$ and $|{01}\rangle$, i.\,e., $|{\underline{1}}\rangle$
has the factor $|{1}\rangle$ in the third position.
\end{enumerate}
}

For the first three cases $\langle{\delta}|{\beta}\rangle=0$ and
$\langle{\gamma}|{\beta}\rangle=0$ implies that $|{\gamma}\rangle$ and
$|{\delta}\rangle$ are linearly dependent or $|{\beta}\rangle =0$. Both
results in a factorization of the code. Thus, for all cases the code
can be factored and thus reduced to a code of length two which
contradicts Theorem~4.

\section{Quantum BCH (${{\cal Q}{\rm BCH}}$) Codes}\label{QBCH}

In principle every quantum error--correcting code applies for the
quantum erasure channel since a $t$ error--correcting code is a $2t$
erasure--correcting code. But even for classical codes, error
correction is a hard task \cite{BMvT78}. The same is true for the
correction of erasures.

But for some codes there are efficient algorithms to correct erasures
and errors. Using the algorithm of Berlekamp and Massey \cite{Mas69}
for decoding binary BCH codes with designed distance ${\rm d_{BCH}}$,
$\nu$ erasures and $t$ errors can be corrected provided that
$\nu+2t<{\rm d_{BCH}}$.

In this section we present a construction of quantum error--correcting
codes based on certain binary BCH codes that can be decoded
efficiently using the algorithm of Berlekamp and Massey.

In a recent preprint \cite{CRSS96} the term {\em quantum BCH code} is
used for codes derived from BCH codes over $GF(4)$. This definition is
more general than ours since every cyclic code over $GF(2)$ is a
subcode of a cyclic code over $GF(4)$. But a BCH code over $GF(4)$
need not be a binary BCH code and thus correction of erasures for the
codes defined in \cite{CRSS96} is not straightforward.

The construction of quantum codes from classical codes is based on the
following theorem \cite{Ste96}:

\noindent{\bf Theorem 6}
{\em Given two classical binary error--correcting codes ${\cal
C}_1=[N,K_1,d_1]$ and ${\cal C}_2=[N,K_2,d_2]$ such that ${\cal C}_1$
contains the dual of ${\cal C}_2$, i.\,e., ${\cal C}^{\perp}_2\le{\cal
C}_1$, a quantum error--correcting code ${\cal
QC}=[[N,K_1-(N-K_2),\min(d_1,d_2)]]$ exists.
}

Here ${\cal C}=[N,K,d]$ denotes a classical binary linear
error--correcting code of length $N$, dimension $K$, and minimum
distance $d$; ${\cal QC}=[[N,K,d]]$ denotes a quantum
error--correcting code with $N$ qubits that encodes $K$ qubits and
allows correction of arbitrary errors of at least $t<d/2$
qubits. Decoding of ${\cal QC}$ is based on (classical) decoding
algorithms for ${\cal C}_1$ and ${\cal C}_2$.

We consider the special case where ${\cal C}_1={\cal C}_2={\cal
C}$. Then ${\cal C}^{\perp}\le{\cal C}$ is required, i.\,e., ${\cal
C}^{\perp}$ has to be a weakly self dual code and an efficient
decoding algorithm for ${\cal C}$ is needed.

For the construction of quantum BCH codes, ${\cal C}$ is chosen to be
a binary BCH code with ${\cal C}^{\perp}$ weakly self dual.

\noindent{\bf Definition 1 (Quantum BCH codes)} \\
{\em
Let ${\cal C}$ be a binary BCH code with ${\cal C}^{\perp}$ weakly
self dual. The states of the quantum BCH code ${{\cal Q}{\rm BCH}}$
code are given (up to normalization) by
$$
|{\psi_{\bbox{v}}}\rangle =
  \sum_{\bbox{c}\in{\cal C}^{\perp}}
|{\bbox{c}+\bbox{v}}\rangle 
  \qquad\mbox{for $\bbox{v}\in{\cal C}/{\cal C}^{\perp}$.}
$$
}

In the remainder of this section we show how to construct the BCH
codes needed for ${{\cal Q}{\rm BCH}}$ codes. First we recall some
properties of BCH codes (for proofs and details see for example
\cite{MS77}).

A cyclic code of length $N$ is defined by the set of roots of its
generator polynomial. The roots are distinct powers of a primitive
$N$--th root $\alpha$. Equivalently, the code corresponds to the set
of exponents of the roots of its generator polynomial, the {\em
defining set} ${\cal I}_{\cal C}$. For binary cyclic codes the
defining set is a union of cyclotomic cosets $C_{i}:=\{i\,2^k \bmod N:
k=0,1,2,\ldots\}$.  For the construction of a binary BCH code with
designed distance ${\rm d_{BCH}}$, ${\cal I}_{\cal C}$ is chosen as
${\cal I}_{\cal C}=C_{b}\cup C_{b+1}\cup\ldots\cup C_{b+{\rm
d_{BCH}}-2},$ i.\,e., the union of cyclotomic cosets of ${\rm
d_{BCH}}-1$ consecutive numbers.

The defining set ${\cal I}_{{\cal C}^{\perp}}$ of the dual code ${\cal
C}^{\perp}$ can be computed from that of the code in the following
manner:
$$
{\cal I}_{{\cal C}^{\perp}}
   =\bigcup_{i\in\overline{\cal I}_{\cal C}} C_{-i},
$$
where $\overline{\cal I}_{\cal C}=\{0,\ldots,N-1\}\backslash{{\cal
I}_{\cal C}}$.  A cyclic code ${\cal C}$ is weakly self dual if and
only if the defining set ${\cal I}_{\cal C}$ contains that of its
dual, i.\,e., ${\cal I}_{{\cal C}^{\perp}}\subseteq{\cal I}_{\cal C}$
or, equivalently, ${\cal I}_{{\cal C}^{\perp}}\cap\overline{\cal
I}_{\cal C}=\emptyset$.

For the ${{\cal Q}{\rm BCH}}$ codes, a binary BCH code ${\cal C}$ with
${\cal C}^{\perp}$ weakly self dual is needed. Therefore, the
condition for ${\cal C}$ is
$$
\bigcup_{i\in{\cal I}_{\cal C}}C_i=
{\cal I}_{\cal C}
\subseteq
{\cal I}_{{\cal C}^{\perp}}
      =\bigcup_{j\in\overline{\cal I}_{\cal C}}C_{-j}
      =\bigcup_{j\notin{\cal I}_{\cal C}}C_{-j}.
$$
Thus, ${\cal I}_{\cal C}$ must not contain both $C_i$ and
$C_{-i}$. Especially, ${\cal I}_{\cal C}$ must not contain cyclotomic
cosets with $C_i=C_{-i}$.

The following lemma summarizes the preceding.

\noindent{\bf Lemma 7 (BCH codes for ${{\cal Q}{\rm BCH}}$ codes)}\\
{\em
Let ${\cal C}$ be a binary BCH code of length $N$ and defining set
${\cal I}_{\cal C}$ such that 
$$
\forall i: \left(
        i \in{\cal I}_{\cal C} \Longrightarrow 
	    (-i \bmod N)\notin{\cal I}_{\cal C}
        \right).
$$
Then the dual code ${\cal C}^{\perp}$ is weakly self dual and a
${{\cal Q}{\rm BCH}}$ code can be constructed.
}

\section{Conclusions}

We conclude by noting that finding efficient codes for restricted
error models is relevant for proof--of--principle demonstrations of
quantum error correction in the near future. The first prototype
quantum computers will presumably have only a few qubits and will not
be powerful enough to implement the most general error correction
schemes.  For example, a simplified demonstration of quantum error
correction could consist in deliberately inducing an error in a known
qubit. In this case the QEC error model applies.

\section{Acknowledgements}

The authors acknowledge fruitful discussions with Peter Zoller and
helpful comments from David DiVincenzo and the referee. T. P. is
supported by the {\it Austrian Science Foundation} under grant
S06514PHY.\@ This research was supported in part by the {\it National
Science Foundation} under grant no.~PHY94-07194.

\end{document}